\documentclass{appolb}
\usepackage{epsfig}

\def\lsim{\raise0.3ex\hbox{$<$\kern-0.75em\raise-1.1ex\hbox{$\sim$}}}
\def\gsim{\raise0.3ex\hbox{$>$\kern-0.75em\raise-1.1ex\hbox{$\sim$}}}
\def\noi{\noindent}
\def\nn{\nonumber}
\def\bea{\begin{eqnarray}}  \def\eea{\end{eqnarray}}
\def\beq{\begin{equation}}   \def\eeq{\end{equation}}

\def\beeq{\begin{eqnarray}} \def\eeeq{\end{eqnarray}}

\def\runhead{}

\begin{document}

\title{Mechanisms of multiparticle production in heavy ion collisions
at high energy\footnote{To be published in a special issue of ``Acta Physica
Polonica'' in homage to Jan Kwiecinski}}

\author{\bf A. Capella\address{
Laboratoire de Physique Th\'eorique\\ Unit\'e Mixte de
Recherche UMR n$^{\circ}$ 8627 - CNRS\\
Universit\'e de Paris XI, B\^atiment 210, F-91405 Orsay Cedex,
France}}

\maketitle

\begin{abstract}
In the framework of a microscopic string model inclusive
charged particle distribution and baryon and antibaryon production
are described. The
emphasis is put on high energies (RHIC) where shadowing
corrections play a crucial role. Some recent developments on $J/\psi$
suppression at CERN-SPS are also discussed. Possible consequences for
the crucial issue of thermal equilibration of the produced system are
considered. \end{abstract}

\section{Introduction}

This work is a continuation of the one presented in ref \cite{1r}, where I
discussed multiparticle production at CERN-SPS energies. The present article is
mainly concerned with higher energies where the efffects of shadowing play a
very important role. In the framework of the model presented below,
the shadowing corrections
can be computed from high mass diffraction practically without any new free
parameter. When the effects of shadowing are taken into account, the model
describes the inclusive charged particle production at RHIC as a function of
centrality \cite{2r}. A comparison with the results obtained \cite{3r,4r} in
the framework of the saturation model is also presented. \par

Another new development concerns net baryon production (stopping). It is shown
that both SPS and RHIC data can be described with the same mechanism (and the
same values of the parameters) used in $pp$ interactions. This indicates that
there is no evidence for an ``anomalous'' stopping in the heavy ion data
\cite{5r}. \par

As already observed at CERN-SPS, rare processes like strange and
multistrange baryon and antibaryon production, can only be described
with the introduction of some final state interaction between the
produced particles (comovers interaction) \cite{1r}. It turns out,
however, that the interaction cross-sections required to describe the
data are comparatively small (a few tenths of a mb) and, in view of the
shortness of the interaction time ($5 \div 7$~fm) it seem quite
improbable that the system can reach thermal equilibrium. In a recent
development \cite{6r} reported below, we show that the same formalism
of final state interaction used at CERN-SPS can describe RHIC data with
the same values of the parameters. Predictions for $\Xi$ and
$\overline{\Xi}$ production have been confirmed by recent STAR data.
Predictions for $\Omega$ and $\overline{\Omega}$ are also given.\par

Finally, we analyze the new NA50 data on $J/\psi$ suppression at
CERN-SPS in the comovers approach \cite{7r} \cite{8r} and discuss
expectations at RHIC.

\section{The model}

\subsection{Hadron-Hadron Interactions}

The Dual Parton Model (DPM) \cite{9r} and the Quark Gluon String Model (QGSM)
\cite{10r} are closely related dynamical models of soft hadronic interactions.
They are based on the large-$N$ expansion of non-perturbative QCD
[11-13] and on
Gribov's Reggeon Field Theory (RFT) \cite{14r}. Their main aim is to determine
the mechanism of multiparticle production in hadronic and nuclear interactions.
The basic mechanism is well known in $e^+e^-$ annihilation (Fig. 1). Here the
$e^+e^-$ converts into a virtual photon, which decays into a $q\overline{q}$
pair. In the rest system of the virtual photon the quark (with colour
3) and the
antiquark (colour $\overline{3}$) separate from each other producing one string
(or chain) of hadrons, i.e. two back-to-back jets. Processes of this type are
called one-string processes. \par

In hadron-hadron interactions, a one-string mechanism is also possible but only
in some cases, namely when the projectile contains an antiquark (quark) of the
same type than a quark (antiquark) of the target, which can annihilate with
each other in their interaction. For instance in $\pi^+p$, the $\overline{d}$
of $\pi^+$ can annihilate with the $d$ of $p$ and a single string is stretched
between the $u$ of $\pi^+$ (colour 3) and a diquark $uu$ of $p$ (colour
$\overline{3}$). This mechanism is also possible in $\overline{p}p$
interactions (Fig. 2) but not in $pp$. This already indicates that it cannot
give the dominant contribution at high energy. Indeed, when taking
the square of the diagram
of Fig. 2 (in the sense of unitarity) we obtain a planar graph, which is the
dominant one according to the large-$N$ expansion. However, this only means
that this graph has the strongest coupling. Since flavour quantum numbers are
exchanged between projectile and target, this graph gives a contribution to the
total cross-section that decreases as an inverse power of $s$ ($1/\sqrt{s})$. A
decrease with $s$ is always associated with flavor exchange. For instance, the
charge exchange $\pi^-p \to \pi^0n$ cross-section also decreases as
$1/\sqrt{s}$. Actually, the diagram in Fig.~2 corresponds to the exchange
of a secondary Reggeon, with intercept close to 1/2. \par

In order to prevent the exchange of flavour between projectile and
target, the $\overline{d}$ and $d$ have to stay, respectively, in the
projectile and target hemispheres. Since they are coloured, they must
hadronize stretching a second string of type $\overline{d}$-$d$. We
obtain in this way a two-string diagram (Figs. 3-5)). Taking the
square of this diagram, we obtain a graph with the topology of a
cylinder (Fig. 6). It turns out that this is the simplest topology one
can construct which does not vanish as $s\to \infty$ due to flavour
exchange. Therefore, we obtain in this way the dominant graph for
hadron-hadron scattering at high energy. The diagram in Fig. 6 is
called a Pomeron (P) and the graphs in Figs. 3-5 a cut Pomeron. Its
order in the large-$N$ expansion is $1/N^2$ [12-13]. Note that due to
energy conservation the longitudinal momentum fractions taken by the
two systems at the string ends have to add up to unity.\par

There are also higher order diagrams (in the sense of the large-$N$
expansion) with 4, 6, 8 strings which give non-vanishing contributions
at high energy. An example of the next-to-leading graph for $pp$
interactions is shown in Fig. 7. It contains four strings -- the two
extra strings are stretched between sea quarks and antiquarks. The
square of this graph corresponds to a $PP$ cut and has the topology of
a cylinder with a handle. Its order in the large-$N$ expansion is
$1/N^4$. The one with six strings corresponds to a $PPP$ cut and to the
topology of a cylinder with two handles (order $1/N^6$), etc. \par

The single particle inclusive spectrum is then given by \cite{9r} \bea
\label{1e} &&{dN^{pp} \over dy}(y) = \sum_n {1 \over \sigma_n} \sum_n
\sigma_n \left ( N_n^{qq-q_v}(y) + N_n^{q_v-qq}(y) + (2n-2)
N_n^{q_s-\overline{q}_s}(y) \right ) \nn \\ &&\simeq N_k^{qq-q_v}(y) +
N_k^{q_v-qq}(y) + (2k-2) N_k^{q_s-\overline{q}_s}(y) \eea

\noi where $k = \sum\limits_n n \sigma_n/\sum\limits_n \sigma_n$ is the
average number of inelastic collisions. Note that each term consists
of $2n$ strings,
i.e. two strings per inelastic collisions. Two of these strings, of
type $qq$-$q$, contain the diquarks of the colliding protons. All other
strings are of type $q$-$\overline{q}$.\par

The weights $\sigma_n$ of the different graphs, i.e. their contribution
to the total cross-section, cannot be computed in the large-$N$
expansion. However, it has been shown \cite{15r} that there is a
one-to-one correspondence between the various graphs in the large-$N$
expansion and those in perturbative Reggeon Field Theory \cite{14r}. We
use the weights obtained from the latter -- with the parameters
determined from a fit to total and elastic cross-sections \cite{9r},
\cite{10r}. At SPS energies we get $k = 1.4$ and at RHIC $k=2$ at
$\sqrt{s} = 130$~GeV and $k = 2.2$ at $\sqrt{s} = 200$~GeV
\cite{2r}.\par

The hadronic spectra of the individual strings $N(y)$ are obtained from
convolutions of momentum distribution functions, giving the probability to
find a given constituent (valence quark, sea quark of diquark) in the
projectile or in the target, with the corresponding fragmentation functions.
The dependence of $N(y)$ on the number of collisions appears via the former.
It is a result of energy conservation. (The larger the number of strings, the
smaller the average invariant mass of each one). \par

Momentum distribution and fragmentation functions are largely
determined from known Regge intercepts \cite{9r} \cite{10r}. The
momentum distribution function of a valence quark in a hadron behaves
as $1/\sqrt{x}$. As in the parton model, this behaviour results from
the intercept 1/2 of a Reggeon trajectory. Thus, in average, the
valence quark in a proton is slow and the diquark is fast due to energy
conservation. Both momentum distribution and fragmentation functions
are assumed to be universal, i.e. the same in all hadronic and nuclear
interactions. This property gives to the model a great predictive
power. Finally, individual strings are assumed to be independent. In
this way, the hadronic spectra of a given graph are obtained by adding
up the corresponding ones for the individual strings. This leads to a
picture, in which, for any individual graph, particles are produced
with only short-range (in rapidity) correlations. Long-range
correlations (and a broadening of the multiplicity distributions) are
due to fluctuations in the number of strings, i.e. to the superposition
of different graphs with their corresponding weights. This gives a
simple and successful description of the data in hadron-hadron and
hadron-nucleus interactions \cite{9r} \cite{10r} \cite{16r}.

\subsection{Nucleus-Nucleus Interactions}

The generalization of Eq. (\ref{1e}) to nucleus-nucleus collisions is rather
straighforward. For simplicity let us consider the case of $AA$ collisions and
let $n_A$ and $n$ be the average number of participants of each nucleus and
the average number of binary $NN$ collisions, respectively. At fixed impact
parameter $b$, we have \cite{17r}
\bea
\label{2e}
&&{dN^{AA} \over dy}(b) = n_A(b) \left [ N_{\mu (b)}^{qq-q_v}(y) +
N_{\mu (b)}^{q_v-qq}(y) + (2 k - 2) N_{\mu (b)}^{q_s-\overline{q}_s}
\right ] \nn \\
&&+ (n(b) - n_A(b)) \ 2 k \ N_{\mu (b)}^{q_s-\overline{q}_s}(y)
\eea

\noi $n_A(b)$ and $n(b)$ are computed from the standard formulae in the
Glauber model. The physical meaning of Eq. (\ref{2e}) is quite obvious.
The expression in brackets corresponds to a $NN$ collision. Since $n_A$
nucleons of each nucleus participate in the collision, this expression
has to be multiplied by $n_A$. Note that in Eq. (\ref{1e}) the average
number of collisions is $k$ -- and the number of strings $2k$. In the
present case the total average number of collisions is $kn$ -- and the
number of strings $2kn$. The second term in Eq. (\ref{2e}) is precisely
needed in order to have the total number of strings required by the
model. Note that there are $2n_A$ strings involving the valence quarks
and diquarks of the participating nucleons. The remaining strings are
necessarily stretched between sea quarks and antiquarks. The value of
$\mu (b)$ is given by $\mu (b) = k \nu (b)$ with $\nu (b) =
n(b)/n_A(b)$ $\mu (b)$ represents the total average number of
inelastic collisions suffered by each nucleon (for more details see
Section 5). \par

We see from eq. (\ref{2e}) that $dN^{AA}/dy$ is obtained as a linear
combination of the average number of participants and of binary
collisions. The coefficients are determined within the model and depend
on the impact parameter via $\mu (b)$. Note that the presence of a term
proportional to the number of binary collisions is a general feature of
RFT and is not related to minijet production. \par

As discussed in Section 2.1 the average invariant mass of a string
containing a diquark at one end is larger than the one of a
$q$-$\overline{q}$ string since the average momentum fraction taken by
a diquark is larger than that of quark. It turns out that the same is
true for the central plateau, i.e.~: $N^{qq-q}(y^* \sim 0) >
N^{q-\overline{q}}(y^* \sim 0)$. Let us now consider two limiting
cases~: \beq
\label{3e}
{\rm If} \ N^{q_s-\overline{q}_s}(y^* \sim 0) \ll N^{qq-q_v} (y^* \sim 0)\ , \
{\rm then}\ {dN^{AA} \over dy}(y^* \sim 0) \sim n_A \sim A^1 \eeq

\beq
\label{4e}
{\rm If} \ N^{q_s-\overline{q}_s}(y^* \sim 0) \sim N^{qq-q_v} (y^* \sim 0)\ , \
{\rm then}\ {dN^{AA} \over dy}(y^* \sim 0) \sim n \sim A^{4/3} \ .
\eeq

\noi In the first case we obtain a proportionality in the number of
participants $n_A$ whereas in the second case we obtain a proportionality in
the number of binary collisions. Since $dN^{AA}/dy \equiv (1/\sigma_{AA})
d\sigma^{AA}/dy$, the latter result implies that $d\sigma^{AA}/dy \sim A^2$,
i.e. all unitarity corrections cancel and we obtain the same result as in the
impulse approximation (Born term only). This result is known as the
Abramovsky-Gribov-Kancheli (AGK) cancellation and is valid for a general class
of models which includes the Glauber and eikonal ones. It implies that, for the
inclusive cross-section, soft and hard processes have the same $A$-dependence.
However, the AGK cancellation is violated by diagrams related to the
diffraction production of large-mass states -- the so-called triple Pomeron or
enhanced diagrams. These diagrams give rise to shadowing corrections as
discussed below. Their effect is very important in nuclear collisions
since they
are enhanced by $A^{1/3}$ factors.

\subsection{Shadowing corrections}

In Appendix A, we discuss the physical content of the AGK cutting rules
and their practical realization in the probabilistic Glauber model. It
is shown there that multiple scattering diagrams, resulting from the
$s$-channel iteration of the Born term, give non-vanishing
contributions to the total cross-sections (shadowing). However, in the
case of the single particle inclusive cross-section, these
contributions cancel identically (AGK cancellation), provided the
measured particle has been produced in an inelastic interaction (cut
Pomeron). If, on the contrary, the trigger particle is produced in the
vertex function (blob) of the multiple scattering diagram, one obtains
the same shadowing effects than in the total cross-section. This is the
physical origin of the AGK violations present in DPM (see section 2.2).
It is clear that if the blob has a small extension in rapidity,
production from the blob will mainly contribute to the fragmentation
region. Therefore, at mid-rapidities, and sufficient large energy, the
AGK cancellation will be valid. \par

Let us consider next the contribution to the total cross-section
resulting from the
diffractive production of large mass states. Clearly, this is
equivalent to an increase
of the rapidity extension of the blob -- which, in this case, can
cover the mid-rapidity
region. Therefore, shadowing corrections to the single particle
cross-section will be
present in this case, provided the measured particle is part of the
diffractively
produced system. As shown in Appendix A, the shadowing correction is
just given by the
diffractive cross-section with negative sign. (This is exactly true
only for purely
imaginary amplitudes). The theoretical expression of the diffractive
cross-section is
well-known. An important part is given by the triple-Pomeron term. It
has also been
measured experimentally and, thus, the shadowing corrections can be
computed with no
free parameters.\par

Considering for simplicity only the contribution of the triple
Pomeron term, the effect
of the shadowing corrections is obtained \cite{18r,2r} by multiplying
Eq. (\ref{2e}) by

\beq
\label{5e}
R_{AB}(b) = {\int d^2s \ f_A(s)\ f_B(b-s) \over T_{AB}(b)}
\eeq

\noi where

\beq
\label{6e}
f_A(b) = {T_A(b) \over 1 + A\ F(s) \ T_A(b)} \ .
\eeq

\noi Here the function $F(s)$ is given by the integral of the ratio
of the triple Pomeron
cross-section $d^2\sigma^{PPP}/dy dt$ at $t = 0$ to the single Pomeron exchange
cross-section $\sigma_p(s)$~:

\beq
\label{7e}
\left . F(s) = 4 \pi \int_{y_{min}}^{y_{max}} dy \ {1 \over \sigma_p(s)} \ {d^2
\sigma^{PPP} \over dy\ dt} \right |_{t=0} = C \left [ \exp \left (
\Delta y_{max}\right )
- \exp \left ( \Delta y_{min}\right ) \right ] \eeq

\noi with $y = \ell n (s/M^2)$, where $M^2$ is the squared mass of
the diffractive
system. For a particle produced at $y_{cm} = 0$, $y_{max} = {1 \over 2} \ell n
(s/m_T)^2$ and $y_{min} = \ell n (R_A m_N/\sqrt{3})$. $\Delta =
\alpha_P(0) - 1 = 0.13$
and $C$ is a constant proportional to the triple Pomeron coupling. $R_A$
is the nuclear radius, $T_A(b)$ the nuclear profile function and
$T_{AB}(b) = \int
d^2s T_A(s) T_B(b-s)$. \par

Eqs. (\ref{5e}) to (\ref{7e}) can be derived only when the triple Pomeron
coupling is small and, thus,  the second term in the denominator of
(\ref{5e}) is small
compared to the first one. In this case, we have $[1 + AF(s)
T_A(b)]^{-1} \sim 1 -
AF(s) T_A(b)$, and only the contribution of the triple Pomeron graph
is involved in the
shadowing. In general higher order rescatterings are also present.
They are model
dependent. The denominator in Eq. (\ref{5e}) correspond to the sum of
all ``fan''
diagrams with Pomeron branchings (generalized Schwimmer model \cite{19r}).\par

It is interesting to study the $A$-dependence of the shadowing
corrections in the limit
of large triple Pomeron coupling (when the first term in the
denominator of (\ref{6e})
can be neglected). In this case we find $R_{AA} \sim A^{-2/3}$, i.e.
the $A^{4/3}$
behaviour resulting from the AGK cancellation is reduced to
$A^{2/3}$. This limit was
considered by Kancheli many years ago \cite{20r}. \par

Note that shadowing corrections to inclusive spectra are not especific to soft
processes. The triple Pomeron terms described above are also
responsible for shadowing
in hard processes.

\section{Charged Particle Multiplicities}

\subsection{Low $p_{T}$}

At SPS energies the limit given by Eq. (\ref{4e}) is not reached, and
Eq. (\ref{2e}) leads to an $A$ dependence of $dN^{AA}/dy$ at $y^* \sim
0$ in $A^{\alpha}$ with $\alpha$ only slightly above unity. ($\alpha
\sim 1.08$ between 2 and 370 participants). On the other hand,
shadowing corrections are small due to phase space limitations
($y_{max} \sim y_{min}$ in Eq. (\ref{7e})). The results \cite{2r} for $Pb$ $Pb$
collisions at $\sqrt{s} = 17.3$~GeV are shown in Fig.~8. We see that
both the absolute values and the centrality dependence are well
reproduced. When the energy increases, Eq. (\ref{4e}) shows that the
value of $\alpha$ should increase towards $4/3$, in the absence of
shadowing corrections. However, the effect of the latter is
increasingly important and, as a result, the value of $\alpha$ varies
little with $s$. At $\sqrt{s} = 130$~GeV, without shadowing corrections
the $A$-dependence is $A^{\alpha}$, with $\alpha \sim 1.27$ in the same
range of $n_{part}$ -- a value which is not far from the maximal one,
$\alpha = 4/3$ from Eq. (\ref{4e}). With the shadowing corrections the
$A$-dependence is much weaker (lower line of the shaded area in
Fig.~9) \cite{2r}. The $A$-dependence is now given by $A^{\alpha}$ with $\alpha
\sim 1.13$ -- always in the range of $n_{part}$ from 2 to 370. As we
see, the increase of $\alpha$ from SPS to RHIC energies is rather
small. This value of $\alpha$ is predicted to change very little
between RHIC and LHC, where $\alpha \approx 1.1$. For, the increase
from $\alpha \sim 1.27$ to $\alpha \sim 4/3$ obtained in the absence of
shadowing is compensated by an increase in the strength of the
shadowing corrections, leaving the effective value of $\alpha$
practically unchanged. \par

\subsection{Large $p_T$}

Let us define the ratio

\beq \label{8e} R_{AA}(b) = {d^2N^{AA}/dy\ dp_T \over n(b) \
d^2N^{pp}/dy\ dp_T} \ . \eeq

\noi For central $Au$ $Au$ collisions at $y^* \sim 0$, we find $R_{AA}
= A^{1.13-4/3} = 0.34$ when numerator and denominator are integrated
over $p_T$. Clearly this value corresponds to small values of $p_T$
which give the dominant contribution to $dN/dy$. This result is in
agreement with the measured values of $R_{AA}$ at $p_T \sim <p_T>$
\cite{21r}. This was to be expected from the results presented in
Fig.~9. It is interesting that these data, as well as PHENIX ones
\cite{22r} show approximately the same value of $R_{AA}$ at large
$p_T$.\footnote{I would like to thank N. Armesto, K.
Boreskov, Y. Dokshitzer, A. Kaidalov, O. Kancheli, A. Krzywicki and
D. Schiff for discussions on this subject.} \par

More precisely, the data show a small increase of $R_{AA}$ at moderate
$p_T$ and, at large $p_T$, they show a scaling in the number of
participants -- rather than in the number of binary collisions. Such a
result is expected in the present approach. Indeed, at large $p_T$ the
shadowing corrections strongly decrease due to the increase of $m_T$ in
$y_{max}$ (eq. (\ref{7e})). However, the larger threshold at large $p_T$
affects mostly the $q$-$\overline{q}$ strings -- which have a smaller
invariant mass than the $qq$-$q$ ones. Thus, with increasing $p_T$ we
approach the limit in eq. (\ref{3e}) leading to a scaling in the number
of participants. The small increase of $R_{AA}$ at moderate $p_T$ is
probably due to the Cronin effect or to a combination of this effect
and jet quenching. Recently, it has been shown \cite{23r} that the
Cronin effect
is considerably smaller at RHIC and LHC energies than at SPS ones due to
the change with energy of the coherence length.

\section{Comparison with the saturation model}

In the saturation model, the $A$-dependence of charged particle
inclusive spectrum in
the saturation regime (i.e. $\Lambda_{QCD} \ll p_T < Q_s$, where
$Q_s$ is the saturation
scale) is given by \cite{3r,4r}

\beq
\label{9e}
{dN \over dy \ d^2p_T} \sim {A^{2/3} \over \alpha_s (Q_s)} \ .
\eeq

\noi Comparison with previous section results indicates that, apart
from logarithmic factors due
to $\alpha_s(Q_s)$, we obtain the same result as in RFT with MAXIMAL
shadowing. As
discussed in the previous section, this result is in violent
disagreement with RHIC
data. \par

The question is then how a reasonable description of the data has
been obtained in
\cite{3r}. The answer is the following. The authors have considered
$dN/dy$ rather than
$dN/dyd^2p_T$. By integrating over $d^2p_T$ up to $Q_s$ and assuming
a $p_T$-broadening
corresponding to $Q_s^2 \sim A^{1/3}$ they have gained one power of $A^{1/3}$.
Furthermore, the factor $\alpha_s^{-1} \sim \log A^{1/3}$ increases
the effective power
slightly above $A^1$ reaching agreement with experiment. The problem with this
explanation is that a $p_T$-broadening in $A^{1/3}$ is much larger
than the one seen in
the data (which is of the order of 30~\% between peripheral and
central collisions). In
order to describe it a parametrization of the saturation scale has
been introduced in
\cite{4r} of the type~:

\beq
\label{10e}
\left [ 0.61 + 0.39 \left ( {n_{part}(b) \over n_{part-max}}\right
)^{1/3} \right ]
p^2_{so} \ .
\eeq

\noi With this definition, $p_{so}$ is the value of the saturation
scale for the most
central collisions (corresponding to $n_{part-max} = 347$
participants). With such an
expression of the saturation scale, the $A$-dependence of ${dN \over
dy}$ is the same
as that of $dN/dyd^2p_T$ within 30~\% and, with the mild $A$-dependence of
$\alpha_s^{-1}$ used before, it is not possible to describe the data.
In view of that
the authors use instead the following expression

\beq
\label{11e}
\left . \alpha_s^{-1} \sim \log  \left [ \left ( 0.61 + 0.39 \left (
{n_{part}(b) \over
n_{part-max}}\right )^{1/3} \right ) \right / \mu^2 \right ]
\eeq

\noi where $\mu^2 = {\Lambda^2_{QCD} \over p_{s,0}^2}$ is chosen to
be 0.6. Note the
fine tuning between the 0.61 in the numerator and 0.6 in the
denominator of (\ref{11e}).
As a consequence, the value of $\alpha_s^{-1}$ for peripheral
collisions is extremely small and $\alpha_s^{-1}$ increases by a
large factor between
peripheral and central collisions. With this fine tuning agreement with experiment is
recovered. Note that for peripheral collisions ($n_{part}(
b) \ll n_{part-max}$) one is
practically sitting on the Landau pole (i.e. the argument of the
$\log$ is very close to
unity). Note also that with $\Lambda_{QCD} = 200$~MeV, the value of
the saturation
scale for the most central collisions is very small (260 MeV). \par

In the saturation model the $A$-dependence of $dN/dyd^2p_T$ grows
larger with increasing $p_T$ and it turns out that the model can
reproduce \cite{24r} the measured values of the ratio $R_{AA}$, Eq.
(\ref{8e}), in the range 1.5 GeV $< p_T < 5$~GeV. However, as discussed
above the model has an $A$-dependence at lower $p_T$ which is too weak.
Thus, it is perhaps not surprising that it has the right
$A$-dependence for larger values of $p_T$ in some range.

The above considerations indicate that saturation is not reached at
RHIC energies. The
considerations in the previous section based on RFT suggest that it
will not be reached
at LHC either.

\section{Nuclear Stopping Revisited}

In $pp$ collisions the net proton ($p$-$\overline{p}$) distribution is
large in the fragmentation regions and has a deep minimum at
mid-rapidities. In contrast to this situation a much flatter
distribution has been observed \cite{25r} in central $Pb$ $Pb$
collisions at CERN-SPS\footnote{Actually, a huge stopping was first
observed at AGS. However, in this case we are in a different regime
(intra-nuclear cascade).}. In view of that, several authors have
claimed that the stopping in heavy ion collisions is anomalous, in the
sense that it cannot be reproduced with the same mechanism (and the
same values of the parameters) used to describe the $pp$ data. In a
recent paper \cite{5r} it has been shown that this claim is not correct.\par

In the model described in previous sections, the net baryon can be
produced directly from the fragmentation of the diquark. Another
possibility is that the diquark splits producing a leading meson in the
first string break-up and the net baryon is produced in a further
break-up. Clearly, in the first case, the net baryon distribution will
be more concentrated in the fragmentation region than in the second
case. The corresponding rapidity distributions are related to the
intercepts of the relevant Regge trajectories, $\alpha_{qq}$ and
$\alpha_q$, respectively, i.e. they are given by $e^{\Delta y(1 -
\alpha )}$. Here
$\Delta y$ is the difference between the rapidity of the produced net
baryon and the maximal one. In the case of the first component, in
order to slow down the net baryon it is necessary to slow down a
diquark. The corresponding Regge trajectory is called baryonium and its
intercept is known experimentally to be $\alpha_{qq} = - 1.5 \pm 0.5$.
For the second component, where a valence quark is slowed down, we take
$\alpha_q = 1/2$ \footnote{There is a third possibility in which the
net-baryon transfer in rapidity takes place without valence quarks
(string junction or gluonic mechanism) with intercept either
$\alpha_{SJ} = 1/2$ \cite{26r} or $\alpha_{SJ} = 1$ \cite{27r}. We find
no evidence for such a component from the existing $pp$ and $AA$ data.
Its smallness could be related to the fact that it produces an extra
string of hadrons and, thus, does not correspond to the dominant
topology.}. \par

In this way we arrive to the following two component model for net
baryon production out
of a single nucleon

\bea
\label{12e}
&&{dN_{\mu}(b) \over dy}(y) = a\ C_{\mu}\ Z_+^{1-\alpha_{q}(0)}(1 - Z_+)^{\mu
(b) - 3/2 + n_{sq}(\alpha_{\rho}(0)-\alpha_{\phi}(0))} \nn \\
&&+ (1 - a) C'_{\mu} \ Z_+^{1- \alpha_{qq} (0)} \times (1 - Z_+)^{\mu
(b) - 3/2 + c + n_{sq}(\alpha_{\rho}(0) - \alpha_{\phi}(0)}
\eea

\noi where $n_{sq}$ is the number of strange quarks in
the hyperon $\alpha_{\rho}(0) = 1/2$ $\alpha_{\phi}(0) = 0$, $Z_+ =
(e^{y-y_{max}})$,
$y_{max}$ is the maximal value of the baryon rapidity and $\mu (b)$
is the average
number of inelastic collisions suffered by the nucleon at fixed
impact parameter $b$ (see section 2.2). The
constants $C_{\mu}$ and $C'_{\mu}$ are obtained from the normalization to unity of each
term. The small $Z$ behaviour is contro
lled by the corresponding intercept. The factor
$(1 - Z_+)^{\mu (b) - 3/2}$ is obtained by requiring that the
$Z$-fractions of all
quarks at the ends of the strings, other than the one in which the
baryon is produced,
go to zero [9-10]. Following conventional Regge rules \cite{28r} an extra
$\alpha_{\rho}(0) = \alpha_{\phi}(0) = 1/2$ is added to the power of
$1 - Z_+$ for each
strange quark in the hyperon. \par

The fraction, $a$, of the DB breaking component is treated as a free
parameter. The same for the parameter $c$ in the DP component -- which
has to be determined from the shape of the (non-diffractive) proton
inclusive cross-section in the baryon fragmentation region. It can be
seen from Eq. (\ref{12e}) that stopping increases with $\mu (b)$, i.e.
with the total number of inelastic collisions suffered by each nucleon.
This effect is present in the two terms of (\ref{12e}) and is a
consequence of energy conservation. The question is whether or not this
``normal'' stopping is sufficient to reproduce the data. In other words
whether or not the data can be described with a universal value of $a$,
i.e. independent of $\mu$ and the same for all reactions.  \par

Eq. (\ref{12e}) gives the total net baryon density,
but it does not allow to determine the relative densities of
different baryon species. In
order to do so we use the simple quark counting rules described in
Appendix B. \par

A good description of the data on the rapidity distribution of $pp
\to p - \overline{p}
+ X$ both at $\sqrt{s} = 17.2$~GeV and $\sqrt{s} = 27.4$~GeV is
obtained from Eq.
(\ref{12e}) with $a = 0.4$, $c = 1$, $\alpha_{q} = 1/2$ and $\alpha_{qq}
= -1$. The results are shown in Table 1 at three different energies,
and compared with
the data. As we see the agreement is reasonable. For comparison with the
nucleus-nucleus results, all values in Table 1 have been scaled by
the number of
participants pairs in central $Pb$ $Pb$ and $Au$ $Au$ collisions
($n_A = 175$). As it
is well known, a pronounced minimum is present at $y^* = 0$. There is also a
substantial decrease of the mid-rapidity yields with increasing
energy. Also, the
mid-rapidity distributions get flatter with increasing energy since
the net proton
peaks are shifted towards the fragmentation regions. \par

It is now possible to compute the corresponding net baryon production
in heavy ion
collisions and to check whether or not the data can be described with
Eq. (\ref{12e})
using the same set of parameters as in $pp$. \par

The results \cite{5r} for net proton $(p-\overline{p})$ and net
baryon $(B - \overline{B})$
production in central $Pb$ $Pb$ collisions at $\sqrt{s} = 17.2$~GeV
and central $Au$
$Au$ collisions at $\sqrt{s} = 130$~GeV are given in Table 2. The
centrality is defined
by the average number of participants -- $n_{part} = 2 n_A = 350$ in
both cases. Experimental
results are given in brackets. \par

The comparison of column 2 with the $pp$ results in Table 1 at the same
energy, shows the well known change in the shape of the rapidity
distribution between $pp$ and central $Pb$ $Pb$ collisions at SPS. The
minimum at $y^* = 0$ is much less pronounced in $Pb$ $Pb$ and the net
proton peaks in the $pp$ fragmentation regions are shifted to $y^* \sim
\pm 1.5$. More interesting are the results in columns 4 and 5 which
contain the predictions for $Au$ $Au$ at RHIC. We see that the shape of
the rapidity distribution is very different from the one at SPS.\par

In conclusion, we have found that ``anomalous'' stopping is not
needed in order to
describe the present data. Related models have been proposed in
[29-31]. The results for heavy ion
collisions are rather similar to the ones obtained from eq.
(\ref{12e}). However, in these models there is some increase in the
size of the second component with the number of inelastic collisions,
i.e. some anomalous stopping is present.

\section{Hyperon-Antihyperon Production}

Strange particle production and, in particular, of multistrange
hyperons has been
proposed as a signal of Quark Gluon Plasma formation. Flavor equilibration is
very efficient in a plasma due to large gluon densities and low
thresholds. An analysis
of the results at SPS in the framework of the present model has been
presented in
\cite{31r}. In the following we concentrate on RHIC results. \par

A general result in DPM is that the ratios $B/h^-$ and
$\overline{B}/h^-$ of baryon and antibaryon yields over negatives
decrease with increasing centralities. This is easy to see from Eq.
(\ref{2e}). The production from $q_s-\overline{q}_s$ strings scales
with the number of binary collisions. These strings have a smaller
(average) invariant mass than the $qq$-$q$ strings and, thus, are more
affected by the thresholds needed for $B\overline{B}$ pair production.
As a consequence, the centrality dependence of $B$ and $\overline{B}$
production will be smaller than the one of negatives. (The same effect
was discussed in Section 3.2 in connection with large $p_T$
production). The effect is rather small at RHIC energies. However, it
is sizable and increases with the mass of the produced baryon. In
contrast with this situation, the data for $\Lambda$'s show no such
decrease and an increase is present for $\Xi$ production. Data on
$\Omega$ production are not yet available. However, SPS data clearly
show a hierachy in the sense that the enhancement of baryon production
increase with the mass (or strange quark content) of the produced
baryon.

The only way out we have found is to give up the assumption of string
independence. Until
now we have assumed that particles produced in different strings are
independent from
each other. In the following we allow for some final state
interactions between comoving
hadrons or partons (see Section 8). We proceed as follows.\par

The hadronic densities obtained in Section 2 are used as initial
conditions in the gain
and loss differential equations which govern final state interactions. In the
conventional derivation \cite{32r} of these equations, one uses
cylindrical space-time
variables and assumes boost invariance. Furthermore, one assumes that
the dilution in
time of the densities is only due to longitudinal
motion\footnote{Transverse expansion
is neglected. The fact that HBT radii are similar at SPS and RHIC and
of the order of
magnitude of the nuclear radii, seems to indicate that this expansion
is not large. The
effect of a small transverse expansion can presumably be taken into
account by a small
change of the final state interactions cross-sections.}, which leads
to a $\tau^{-1}$
dependence on the longitudinal proper time $\tau$. These equations
can be written
\cite{32r} \cite{31r}

\beq
\label{13e}
\tau \ {d\rho_i \over d \tau} = \sum_{k\ell} \sigma_{k\ell} \ \rho_k
\ \rho_{\ell} -
\sum_k \sigma_{ik}\ \rho_i \ \rho_k \ . \eeq

\noi The first term in the r.h.s. of (\ref{13e}) describes the
production (gain) of
particles of type $i$ resulting from the interaction of particles $k$
and $\ell$. The
second term describes the loss of particles of type $i$ due to its
interactions with
particles of type $k$. In Eq. (\ref{13e}) $\rho_i = dN_i/dyd^2s(y,b)$
are the particles
yields per unit rapidity and per unit of transverse area, at fixed
impact parameter.
They can be obtained from the rapidity densities (\ref{2e}) using the
geometry, i.e. the
$s$-dependence of $n_A$ and $n$. The procedure is explained in detail
in \cite{7r} where
the pion fragmentation functions are also given. Those of kaons and
baryons can be found in
\cite{6r}. $\sigma_{k\ell}$ are the corresponding cross-sections
averaged over the
momentum distribution of the colliding particles. \par

Equations (\ref{13e}) have to be integrated from initial time $\tau_0$
to freeze-out time $\tau_f$. They are invariant under the change $\tau
\to c \tau$ and, thus, the result depends only on the ratio
$\tau_f/\tau_0$. We use the inverse proportionality between proper time
and densities and put $\tau_f/\tau_0 = (dN/dyd^2s(b))/\rho_f$. Here the
numerator is given by the DPM particles densities. We take $\rho_f =
[3/\pi R_p^2](dN^-/dy)_{y^* \sim 0} = 2$~fm$^{-2}$, which corresponds
to the density of charged and neutrals per unit rapidity in a $pp$
collisions at $\sqrt{s} = 130$~GeV. This density is about 70 \% larger
than at SPS energies. Since the corresponding increase in the $AA$
density is comparable, the average duration time of the interaction
will be approximately the same at CERN SPS and RHIC -- about 5 to 7
fm.\par

Next, we specify the channels that have been taken into account in
our calculations.
They are

\beq
\label{14e}
\pi N \stackrel{\rightarrow}{\leftarrow} K \Lambda (\Sigma)\ , \quad
\pi \Lambda (\Sigma )
\stackrel{\rightarrow}{\leftarrow} K \Xi \ , \quad \pi \Xi
\stackrel{\rightarrow}{\leftarrow} K \Omega  \eeq

\noi We have also taken into account the strangeness exchange reactions

\beq
\label{15e}
\pi \Lambda (\Sigma ) \stackrel{\rightarrow}{\leftarrow} K N\ , \quad \pi \Xi
\stackrel{\rightarrow}{\leftarrow} K \Lambda (\Sigma ) \ , \quad \pi
\Omega  \stackrel{\rightarrow}{\leftarrow}
K \Xi \eeq

\noi as well as the channels corresponding to (\ref{14e}) and (\ref{15e}) for
antiparticles\footnote{To be precise, of all possible charge
combinations in reactions
(\ref{14e}), we have only kept those involving the annihilation of a light
$q$-$\overline{q}$ pair and production of an $s$-$\overline{s}$ in
the $s$-channel. The
other reactions, involving three quarks in the $t$-channel
intermediate state, have
substantially smaller cross-sections and have been neglected. All
channels involving
$\pi^0$ have been taken with cross-section $\sigma /2$ since only one of the
$u\overline{u}$ and $d\overline{d}$ components of $\pi^0$ can
participate to a given
charge combination. For details see the first paper of \cite{31r}.}.
We have taken
$\sigma_{ik} = \sigma = 0.2$~mb, i.e. a single value for all
reactions in (\ref{14e}) and
(\ref{15e}) -- the same value used in ref. \cite{31r} to describe the
CERN SPS data.
\par

Before discussing the numerical results and the comparison with
experiment let us
examine the qualitative effects of comovers interaction. As explained
in the beginning of
this Section, without final state interactions all ratios $K/h^-$, $B/h^-$
and $\overline{B}/h^-$ decrease with increasing centrality. The final
state interactions (\ref{14e}), (\ref{15e}) lead to a gain of strange
particle
yields.  The reason for this is the following. In the first direct
reaction (\ref{14e}) we
have $\rho_{\pi} > \rho_K$, $\rho_N > \rho_{\Lambda}$, $\rho_{\pi}
\rho_N \gg \rho_K
\rho_{\Lambda}$. The same is true for all direct reaction
(\ref{14e}). In view of that,
the effect of the inverse reactions (\ref{14e}) is small. On the
contrary, in all
reactions (\ref{15e}), the product of densities in the initial and
final state are
comparable and the direct and inverse reactions tend to compensate
with each other.
Baryons with the largest strange quark content, which find themselves
at the end of the
chain of direct reactions (\ref{14e}) and have the smallest yield
before final state
interaction, have the largest enhancement. Moreover, the gain in the
yield of strange
baryons is larger than the one of antibaryons since $\rho_B >
\rho_{\overline{B}}$.
Furthermore, the enhancement of all baryon species increases with
centrality, since the
gain, resulting from the first term in Eq. (\ref{13e}), contains a
product of densities
and thus, increases quadratically with increasing centrality.\par

\subsection{Numerical Results}

All our results refer to mid-rapidities. The calculations have been
performed in the interval $-0.35 < y^* < 0.35$. In Fig.~10a-10d we
show the rapidity
densities of $B$, $\overline{B}$ and $B - \overline{B}$\footnote{In
the numerical
calculations the net baryon yields have been obtained using the
approach in \cite{6r}
and \cite{31r}. This approach is conceptually different from the one
in Section 5 but
the numerical results are similar.} versus $h^- = dN^-/d\eta =
(1/1.17) dN/dy$ and compare
them with available data [33-35]. We would like to stress that the
results for $\Xi$ and
$\overline{\Xi}$ were given \cite{6r} before the data \cite{35r}. This is an important
success of our approach.\par

In first approximation,
the yields of $p$, $\overline{p}$, $\Lambda$ and
$\overline{\Lambda}$ yields over $h^-$ are independent of centrality.
Quantitatively, there is a slight decrease with centrality of
$p/h^-$ and $\overline{p}/h^-$ ratios, a slight increase of $\Lambda
/h^-$ and $\overline{\Lambda}/h^-$ and a much larger increase for
$\Xi$
($\overline{\Xi})/h^-$ and $\Omega$ ($\overline{\Omega})/h^-$. This
is better seen in
Fig.~11a and 11b where we plot the yields of $B$ and $\overline{B}$
per participant
normalized to the same ratio for peripheral collisions versus
$n_{part}$. The enhancement
of $B$ and $\overline{B}$ increases with the number of strange quarks
in the baryon. This
increase is comparable to the one found at SPS between $pA$ and
central $Pb$ $Pb$ collisions, especially for antibaryons. The ratio
$K^-/\pi^-$ increases
by 30~\% in the same centrality range, between 0.11 and 0.14 in
agreement with present
data. The ratios $\overline{B}/B$ have a mild decrease with centrality of about
15~\% for all baryon species -- which is also seen in the data. Our values for
$N^{ch}/N_{max}^{ch} = 1/2$ are~: $\overline{p}/p = 0.69$,
$\overline{\Lambda}/\Lambda =
0.74$, $\overline{\Xi}/\Xi = 0.79$, $\Omega/\overline{\Omega} = .83$,
to be compared with
the measured values \cite{36r}~:

$$\overline{p}/p = 0.63 \pm 0.02 \pm 0.06 \quad , \quad
\overline{\Lambda}/\Lambda = 0.73 \pm 0.03 \quad , \quad
\overline{\Xi}/\Xi = 0.83 \pm
0.03 \pm 0.05 \ .$$

\noi The ratio $K^+/K^- = 1.1$ and has a mild increase with
centrality, a feature also
seen in the data. \par

Note that a single parameter has been adjusted in order to determine
the absolute yields
of $B\overline{B}$ pair production, namely the $\overline{p}$ one --
which has been
adjusted to the experimental $\overline{p}$ value for peripheral
collisions. The yields of all other $B\overline{B}$ pairs has been
determined using the
quark counting rules given in Appendix B. The experimental data in
Fig. 10 are not
corrected for feed-down from weak decays. If these corrections were the same
(in percentage) for all baryon species, our results should be compared with
uncorrected yields. This seems to be the case for $p$, $\overline{p}$,
$\Lambda$ and $\overline{\Lambda}$ where the feed-down corrections
are of the order of 20~\%.
As a consequence, our predictions for $\Xi$, $\overline{\Xi}$, $\Omega$ and
$\overline{\Omega}$ have a 20~\% uncertainty. \par

Although the inverse slopes (``temperature'') have not been discussed
here, let us note
that in DPM they are approximately the same for all baryons and
antibaryons both before and after final state
interaction -- the effect of final state interaction on these slopes
being rather small \cite{37r}.

\section{New $J/\psi$ suppression data and the comovers interpretation}

The NA38-NA50 collaboration have observed a decrease of the ratio of
$J/\psi$ to dimuon (DY) cross-sections with increasing centrality in
$SU$ and $Pb$ $Pb$ collisions. The same phenomenon has been observed in
$pA$ collisions with increasing values of $A$. In this case, it is
interpreted as due to the interaction of the pre-resonant
$c\overline{c}$ pair with the nucleons of the nucleus it meets in its
path (nuclear absorption). As a result of this interaction, the
$c\overline{c}$ pair is modified in such a way that, after interaction,
it has no projection into $J/\psi$ (a $D\overline{D}$ pair is produced
instead). The $J/\psi$ survival probability $S_{abs}$ is well known
(see for instance eq. (7) of \cite{7r}) and depends on a single free
parameter $\sigma_{abs}$, i.e. the absorptive $c\overline{c} - N$ cross
section. \par

The NA50 collaboration has shown that the $J/\psi$ suppression in $Pb$
$Pb$ collisions has an anomalous component, i.e. it cannot be
reproduced using nuclear absorption alone. Two main interpretations have
been proposed~: deconfinement and comovers interaction. The latter
mechanism has been described in Section 6 for strange particle
production. In the case of $J/\psi$ suppression, a single channel is
important namely $c\overline{c}$ (or $J/\psi$) interacting with comoving
hadrons and producing a $D\overline{D}$ pair. In this case, eq.
(\ref{13e}) can be solved analytically. The expression of the survival
probability $S_{co}$ can be found in \cite{7r} (see eq. (8)). It
depends on a free parameter $\sigma_{co}$, i.e. the effective
cross-section for the comovers interaction. \par

Two important sets of new data have been presented recently by the NA50
collaboration on $pA$ \cite{38r} and $Pb$ $Pb$ collisions \cite{39r}.
Before these data were available, the NA50 interpretation of the data
was as follows. The $pA$, $SU$ and peripheral $Pb$ $Pb$ data can be
described with nuclear absorption alone, with $\sigma_{abs} = 6.4 \pm
0.8$~mb. At $E_T \sim 40$~GeV there is a sudden onset of anomalous
suppression with a steady fall off at large $E_T$. However, at variance
with this view, the most peripheral $Pb$ $Pb$ points lied above the
nuclear absorption curve -- which extrapolates $pA$ and $SU$ data.\par

The new $pA$ data indicate a substantially smaller value of the
absorptive cross-section. However, within errors, $pA$ and $SU$ data
can still be described with $\sigma_{abs} = 4.4 \pm 0.5$~mb \cite{38r}.
The new $Pb$ $Pb$ preliminary data, taken in 2000 with a target under
vacuum, are
consistent with previous ones except for the most peripheral ones --
which are now lower and consistent with the nuclear absorption curve
\cite{39r}. In this way, the NA50 interpretation remains valid.
However, the new data lend support to the
interpretation based on comovers interaction. Indeed, due to the
smaller value of $\sigma_{abs}$ there is more room for comovers
interaction (i.e. for anomalous suppression) in
$SU$. \par

Actually, before the new data were available, it has been argued \cite{8r}
that a value of $\sigma_{abs} = 4.5$~mb is also consistent within
errors with the old $pA$ data. Using this value and $\sigma_{co} =
1$~mb it has been possible to describe all available data within the
comovers scenario \cite{7r,8r,40r}. There was, however, a caveat, as pointed
out in \cite{8r}. Indeed, there was a mismatch of about 30~\% between
the absolute normalizations in $SU$ and $Pb$ $Pb$. Actually, the
ratio of the first
normalization to the second one is only $1.04 \pm 0.02$ \cite{38r}.
(This factor takes
into account both the isospin correction in $SU$ and the rescaling in energy).
This mismatch was induced by the high values of the most peripheral
$Pb$ $Pb$ data in the old NA50 data. Indeed, since the relative
contribution of the comovers to $J/\psi$ suppression is larger for
central collisions, the centrality dependence of the $J/\psi$
suppression gets flatter (steeper) with decreasing (increasing) values
of $\sigma_{co}$ (at fixed $\sigma_{abs}$). In order to reproduce the
shape of the $J/\psi$ over $DY$ ratio, from very peripheral to central
collisions, in the old NA50 analysis, a value of $\sigma_{co} = 1$~mb
was required. On the other hand the new data are described with a
smaller value $\sigma_{co} = 0.65$~mb. This decrease of $\sigma_{co}$
leads to a decrease of the absolute normalization, which is now
consistent with the $SU$ one\footnote{It is interesting that
almost the same value of $\sigma_{co}$ ($\sigma_{co} = 0.7$~mb) was
obtained in \cite{41r} from an analysis of $SU$ data and old $Pb$ $Pb$
data (which covered a much smaller centrality range). In \cite{41r} the
absolute normalization in $SU$ and $Pb$ $Pb$ were in good agreement
with each other.}. \par

The results \cite{42r} of the comover interaction model with
$\sigma_{abs} = 4.5$~mb and $\sigma_{co} = 0.65$~mb are presented in
Fig. 12. As in ref. \cite{40r} the steady fall-off of the $J/\psi$ over
$DY$ ratio at large $E_T$ is obtained introducing the $E_T$
fluctuations. The agreement with the new NA50 data \cite{39r} is quite
satisfactory. The absolute normalization is 47. The corresponding one
in $SU$ is 45 in perfect agreement with the expectations discussed
above.\par

It is interesting to note that the data obtained using the $E_T$
calorimeter and the zero-degree calorimeter (ZDC) analysis are
consistent with each other when using the measured $E_T-E_{ZDC}$
correlation. This result was predicted in ref. \cite{8r}.\par

Next, I would like to discuss briefly the expectations for $J/\psi$
supression at RHIC in the comovers interaction model. The calculation
of the survival probability $S_{co}$ is quite safe. Indeed, since
$\sigma_{co}$ is a cross-section near threshold, the same value obtained at
SPS should be used at RHIC. The situation is quite different for
$S_{abs}$. Many authors assume that $\sigma_{abs}$ is the same at RHIC
and at SPS. It has also been suggested that it can be significantly
larger at RHIC. However, it seems plausible that at mid-rapidities,
nuclear absorption at RHIC is small due to the fact that, contrary to
SPS, the $c\overline{c}$ pair is produced outside the colliding nuclei.
It is therefore crucial to have data on $J/\psi$ production in $pA$
interactions at RHIC. If $S_{abs} \sim 1$ the $J/\psi$ suppression at
RHIC and SPS will be comparable since the smallness of the nuclear
absorption will be approximately compensated by the increase of the
comovers suppression -- due to a larger comovers density at RHIC. Very
preliminary data tend to indicate that this is indeed the case. \par

A quantitative analysis of the new NA50 data in the deconfining scenario is still
missing. On the other hand, the centrality dependence of the average $p_T$ of
$J/\psi$ is better described in the comovers approach than in a deconfining
scenario \cite{43r}. At RHIC energies, a small nuclear absorption in $pA$
collisions (i.e. $S_{abs} \sim 1$), would be a very interesting situation in
order to discriminate the comovers interaction model from a deconfining scenario.
Indeed, in the latter, the shape of the centrality dependence would be almost
flat for peripheral collisions (below the deconfining threshold) and would
decrease above the threshold. Such a behaviour would be a clear signal of
deconfinement. On the contrary, in the comovers scenario, the fall-off would be
continuous, from peripheral to central collisions, and determined by the same
value of $\sigma_{co}$ obtained from CERN SPS data.

\section{Conclusions}

Quark Gluon Plasma (QGP) formation is obtained in statistical QCD,
i.e. QCD applied to a
system in thermal equilibrium. Therefore, one of the main issues in
heavy ions physics
is to determine whether or not the produced final state reaches
thermal equilibrium. An
argument in favor of equilibrium is the fact that particle abundances
are well described in
terms of statistical models. However, one should take into account
that statistical
models are also very successful in $pp$ and even $e^+e^-$
interactions. Therefore, it is
important to study whether or not particle abundances can be obtained
in a microscopic
model such as DPM. \par

As a starting point we have assumed that particles produced in
different strings are independent (see Section 2). In this case thermal
equilibrium cannot be reached no matter how large the energy density
is. Indeed, in this case a large energy-density is the result of
piling up a large
number of independent strings. The assumption of independence of
strings works remarkably well in $hh$ and $hA$ interactions \cite{9r}
\cite{10r} -- even in the case of event samples with 5 or 6 times the
average multiplicity -- indicating that no sizable final state
interaction is present in these reactions. In nucleus-nucleus
collisions, we have described charged particle inclusive production and
its centrality dependence. The model exhibits a term proportional to
the number of binary collisions -- which has been seen in the data both
at SPS and RHIC. The presence of such a term is required by unitarity
and is not due to minijets. \par

However, it is clear that in heavy ion collisions, where several
strings occupy a transverse area of 1 fm$^2$, the assumption of string
independence has to break down. This is indeed the case. As we have
seen, some data cannot be described without final state interaction. It
could have happened that this final state interaction is so strong that the
string picture breaks down and becomes totally useless. This does not
seem to be the case. On the contrary, present data can be described
using the particle densities computed in the model as initial
conditions in the gain and loss (transport) equations governing the
final state interaction. The interaction cross-section turns out to be
small (of the order of a few tenths of a mb). Due to this smallness and
to the limited interaction time available, final state interaction has
an important effect only on rare processes, in particular $\Xi$,
$\Omega$ and $J/\psi$ production. The bulk of the final state is not
affected. \par

Of course it is not possible to conclude that thermal equilibrium has
not reached. However, particle abundances not only do not allow to
conclude that it has been reached, but, on the contrary, their centrality
dependence tends to indicate that this is not the case. Let us consider
for instance $p$ and $\overline{p}$ production. In our model, their
yields are practically not affected by final state interaction, i.e.
they are practically the same assuming string independence. Yet, the
model reproduces the data, from very peripheral to very central
interaction. This success would be difficult to understand in a QGP
scenario in which for peripheral collisions (below the critical
density) there is strong, non-equilibrated, $p\overline{p}$
annihilation, which becomes equilibrated for central ones, above the
critical density. More generally, the QGP scenario would be strongly
supported if some kind of threshold would be found in the
strange baryon yields around the critical density value. At SPS energies,
evidence for such a threshold in the $\overline{\Xi}$ yield has been
claimed by the NA57 collaboration \cite{44r}. Unfortunately, these data
only cover a limited range of centrality. In contrast to this situation
the RHIC data explore the whole centrality range from very peripheral
to very central collisions and the centrality dependence of the yields
of $p$, $\Lambda$, $\Xi$ and their antiparticles shows no structure
whatsoever. If the same happens for $\Omega$ and $\overline{\Omega}$
production (as predicted in our approach) the case for QGP formation
from strange baryon enhancement will be rather weak.\par

Finally, it should be stressed that the final state interaction of
comovers in our approach is by no means a trivial hadronic effect.
Indeed, the interaction of comovers starts at the early
times where densities, as computed in DPM, are very large. In this
situation the comovers are not hadrons (there are several of them in
the volume occupied by one hadron, and, moreover, at these early times
hadrons are not yet formed). This is probably the reason
why in our approach the comover interaction cross-sections required to
describe the data are smaller than in a hadron gas model. \par

Before concluding, I would like to say that it is an honor and a pleasure to
contribute to this special issue of Acta Physica Polonica in hommage
to my friend
Jan Kwiecinski. I would also like to acknowledge his important
contributions to the
model presented here, realized during his (too rare) visits to Orsay.
In particular he
played an important role in the generalization of the Dual Parton
Model to heavy
ions collisions \cite{17r} and also in introducing \cite{45r} a
semi-hard component in
the model. \par \vskip 1 truecm

\centerline{\large\bf  Appendix A} \par \vskip 1 truecm

{\it a) Reggeon Field Theory versus Glauber Model} \par \vskip 3 truemm

The reggeon calculus or reggeon field theory (RFT) \cite{14r}
provides a field theoretical
formulation of the eikonal (for $hh$ collisions) or the Glauber (for
$hA$ and $AB$)
models, valid at high energies. The main difference between the RFT
and the Glauber
model is that, at high energies, the coherence length is large and
the whole nucleus
is involved in the interaction. Moreover, due to the space-time
development of the
interaction, when, at high energy a projectile interacts
inelastically with a nucleon
of the nucleus, the formation time of (most of) the produced
particles is larger than
the nuclear size and, thus, particles are produced outside the
nucleus. Therefore,
planar diagrams give a vanishing contribution at high energy. The
relevant diagrams
are non-planar, describing the ``parallel'' interactions of constituents of the
projectile with the target nucleons (in the case of an $hA$
collision). This picture
is in clear contrast with the Glauber model, in which the projectile undergoes
successive (billiard ball type of) collisions with the nucleons of
the target. \par

In spite of these differences, one recovers the Glauber formula in first
approximation. This formula corresponds to the contribution of the
initial state
(on-shell projectile pole) to the various rescattering terms. In RFT
one has, besides
these contributions, also the contributions due to low mass and high
mass diffractive
excitations of the projectile. The latter are very important since,
as we have seen in
Section 2.3, they give rise to shadowing corrections.\par \vskip 5 truemm

{\it b) Cutting Rules} \par \vskip 3 truemm

An important feature of RFT is that it obeys to the so-called AGK
cutting rules \cite{46r}.
These rules allow to relate to each other the different $s$-channel
discontinuities of a
given graph, and also to relate them to the contribution of this
graph to the total
cross-section. In this way, they provide a powerful link between
total cross-section and
multiparticle production. In order to illustrate these rules, let us
consider the case
of an interaction of a hadron $h$ with two different nucleons of the
target nucleus $A$
(with $A-2$ spectators), and let us assume that the object exchanged
in the $t$-channel
of each collision is purely imaginary (Pomeron). \par

Let us consider the cutting by a plane in between the two interactions
(i.e. in between the two Pomerons). We obtain in this way a diffractive
intermediate state containing a large rapidity gap. Let us call $+ 1$
its contribution to $\sigma_{tot}$. From the cutting rules we find that
the inelastic contribution obtained cutting through one of the
interactions (an interference term) has a weight $-2$ relative to the
previous one. Since there are two interactions one can cut through, one
obtains $- 4$. Finally, cutting by a plane through the two interactions
(which is possible since the graph is non-planar) has a relative weight
$+ 2$. This last contribution is also inelastic and has an average
multiplicity which is twice that of the previous one. The total
contribution of this double scattering, to $\sigma_{tot}$ is thus equal
to $+ 1 - 4 + 2 = - 1$, a negative contribution. The total contribution
to the non-diffractive inelastic cross-section is $\sigma_{ND} = - 4 +
2 = - 2$. We see in this way that the (negative) contribution of a
double interaction to $\sigma_{ND}$ is two times larger, in absolute
value that its contribution to $\sigma_{tot}$. In the case of $n$
collisions the corresponding factor is $2^n$. \par

Let us now consider the contribution of a double interaction to the
non-diffractive
single particle inclusive cross-section $d\sigma/dy$. This
contribution is $- 4 + 2
\times 2 = 0$. Indeed, in the case of the cut through the two interactions the
contribution to $d\sigma/dy$ has an extra factor 2 since the triggered
particle can be produced in either of the two interactions. It turns
out that such a
cancellation is true to all orders in the number of interactions. We
obtain in this way
the so-called AGK cancellation. All rescattering corrections of the
Glauber type cancel
identically in $d\sigma/dy$. Only the term with a single interaction
is left -- which
is proportional to $A^1$ in $pA$ interactions.  \par

Note that the crucial ingredient in obtaining the AGK cancellation is
the fact that the triggered particle has been produced in a cut
interaction -- which gives the extra factor 2 for two cut
interactions. The other possibility is that,  the trigger particle is
emitted from the (cut) vertex function (blob). Clearly, in this case
the extra factor 2 is absent and the AGK cancellation is not valid. In
this case the shadowing corrections are the same as in the total
cross-section.\par

The AGK cutting rule described above are quite general. They are
valid in any field
theory in which the vertex functions obey the general properties of
unitarity, crossing
and large $p_T$ damping. The Glauber model is a particular example in
which the AGK
rules are valid. Their derivation in this case is straightforward, as
discussed below.
\par \vskip 5 truemm

{\it c) The Probabilistic Glauber Model and the Cutting Rules} \par \vskip 3
truemm

Let us consider for simplicity $pA$ scattering. The main formula of the
probabilistic Glauber model is the one that gives the cross-section
$\sigma_n$ for $n$ inelastic collisions of the projectile with $n$
nucleons of the target nucleus, at fixed impact parameter $b$~:

$$\sigma_n (b) = {A \choose n} \left ( \sigma_{inel} \ T_A(b) \right
)^n \left ( 1 - \sigma_{inel} \ T_A(b) \right )^{A-n} \eqno({\rm A}.1)$$

\noi where $\sigma_{inel}$ is the proton-nucleon inelastic
cross-section and $T_A(b)$ is
the nuclear profile function. This equation is just the Bernoulli's formula for
composite probabilities. The first factor is a trivial combinatorial factor
corresponding to the different ways of choosing $n$ nucleons out of
$A$. The second one
gives the probability of having $n$ inelastic $pN$ collisions at
given $b$. The third
one is the probability that the remaining $A - n$ nucleons do not interact
inelastically. Let us consider first a term with two collisions both
of which are
inelastic. The corresponding cross-section is $\sigma_{2}^2(b) = {A \choose 2}
(\sigma_{inel} T_A(b))^2$ i.e. a positive term. Let us now consider
the case of two
collisions only one of which is inelastic. The corresponding
(interference) term is
$\sigma_2^1(b)$ obtained from eq. (A.1) by putting $n = 1$ and taking
the second term
in the expansion of the last factor. We get $\sigma_2^1(b) = -
A(A-1)(\sigma_{inel}T_A(b))^2$. We see that $\sigma_2^1(b) = - 2
\sigma_2^2(b)$. Thus,
a rescattering term containing two collisions gives a negative contribution to
$\sigma_{tot}$. \par

Let us now consider the contribution to $d\sigma /dy$. It is given by
$\sigma_2^1(b) + 2 \sigma_2^2(b) = 0$. Indeed, in the case of a
double inelastic
collision, the triggered particle can be emitted in either of them --
hence an extra
factor 2. This is just the AGK cancellation. It is easy to see that
it is valid order
by order in the total number of collisions. This can also be seen as
follows. The
total inelastic cross-section for $pA$ collision in the Glauber model
is given by the
well known expression

$$\sigma_{inel}^{pA} (b) = \sum_{n=1}^A \sigma_n (b) = 1 - \left ( 1
- \sigma_{inel} \
T_A(b) \right )^A \ . \eqno({\rm A}.2)$$

\noi This expression contains a term in $A^1$ (Born term or impulse
approximation). It
also contains contribution from multiple scattering with alternate
signs. Numerically,
it behaves as $A^{\alpha}$ with $\alpha \sim 2/3$. The single
particle inclusive
cross-section is given by

$${d\sigma^{pA} \over dy}(b) \propto \sum_{n=1}^A n\ \sigma_n(b) = A
\ \sigma_{inel}\
T_A(b) \ . \eqno({\rm A}.3)$$

\noi We see that here multiple-scattering contributions cancel
identically and only the Born term is left. As a consequence of this
AGK cancellation the $A$-dependence of $d\sigma/dy$ in $pA$
interactions behaves as $A^1$. In the case of $AB$ collisions it
behaves as $AB$ and $dN^{AB}/dy = (1/\sigma_{AB})
d\sigma^{AB}/dy$ is proportional to the number of binary collisions --
rather than to the number of participants. \par

We see in this way that the AGK rules are trivially satisfied in the
Glauber model. As mentioned in Section A.1 in the Glauber model only the
initial state is present in the vertex function (blob). Thus a secondary
can only be produced in an interaction and the AGK cancellation is
exact. In a general theory with a more complicated vertex function, the
triggered particle may be produced in the blob. As discussed in Section
A.2 this gives rise to a violation of the AGK cancellation -- which is
responsible for the shadowing corrections to the inclusive spectra.
\par \vskip 1 truecm

\centerline{\large\bf  Appendix B} \par \vskip 1 truecm

In order to get the relative densities of each baryon and antibaryon
species we use
simple quark counting rules \cite{6r} \cite{31r}. Denoting the
strangeness suppression factor by
$S/L$ (with $2L+ S = 1$), baryons produced out of three sea quarks
(which is the case for pair
production) are given the relative weights

$$I_3 = 4L^3 : 4L^3 : 12L^2S : 3LS^2 : 3LS^2 : S^3 \eqno({\rm B}.1)$$

\noi for $p$, $n$, $\Lambda + \Sigma$, $\Xi^0$, $\Xi^-$ and $\Omega$,
respectively. The
various coefficients of $I_3$ are obtained from the power expansion
of $(2L + S)^3$.\par

For net baryon production, we have seen in Section 5 that the baryon
can contain
either one or two sea quarks. The first case corresponds to direct
diquark fragmentation
described by the second term of Eq. (\ref{12e}). The second case
corresponds to diquark
splitting, described by the first term of (\ref{12e}). In these two
cases, the relative
densities of each baryon species are respectively given by

$$I_1 = L : L : S \eqno({\rm B}.2)$$

\noi for $p$, $n$ and $\Lambda + \Sigma$, and

$$I_2 = 2L^2 : 2L^2 : 4LS : {1 \over 2} S^2 : {1 \over 2} S^2
\eqno({\rm B}.3)$$

\noi for $p$, $n$, $\Lambda + \Sigma$, $\Xi^0$ and $\Xi^-$. The
various coefficients in
(B.2) and (B.3) are obtained from the power expansion of $(2L + S)$
and $(2L + S)^2$,
respectively.\par

In order to take into account the decay of $\Sigma^*(1385)$ into
$\Lambda \pi$, we
redefine the relative rate of $\Lambda$'s and $\Sigma$'s using the
empirical rule
$\Lambda = 0.6(\Sigma^+ + \Sigma^-$) -- keeping, of course, the total yield of
$\Lambda$'s plus $\Sigma$'s unchanged. In this way the normalization
constants of all
baryon species in pair production are determined from one of them.
This constant,
together with the relative normalization of $K$ and $\pi$, are
determined from the data
for very peripheral collisions. In the calculations we use $S = 0.1$
$(S/L = 0.22)$.

\newpage
\centerline{\bf Table 1} \par \vskip 3 truemm

Calculated values \cite{5r} of the rapidity distribution of $pp \to p
- \overline{p} + X$ at
$\sqrt{s} = 17.2$~GeV and 27.4 GeV ($k = 1.4$) and $\sqrt{s} =
130$~GeV ($k=2$). (In
order to convert $d\sigma/dy$ into $dN/dy$ a value of $\sigma =
30$~mb has been used).
For comparison with the nucleus-nucleus results, all values in this table have
been scaled by $n_A = 175$ -- the number of participant pairs in
central $Pb$ $Pb$ and
$Au$ $Au$ collisions. Data are in brackets.

\begin{center}
\begin{tabular}{cccc}
\hline
$y^*$ &$pp \to p - \overline{p}$ &$pp \to p - \overline{p}$ &$pp \to
p - \overline{p}$ \\
&$\sqrt{s} = 17.2$ GeV &$\sqrt{s} = 27.4$ GeV &$\sqrt{s} = 130$ GeV \\
\hline
0 &9.2 &6.5 &3.6 \\
& &$[6.3 \pm 0.9 ]$ & \\
1 &15.0 &9.3 &4.2 \\
&$[16.1 \pm 1.8]$ &$[9.6 \pm 0.9]$ & \\
1.5 &25.8 &14.6 &5.1 \\
&$[24.1 \pm 1.4]$ &$[15.4 \pm 0.9]$ & \\
2 &47.1 &26.2 &6.8 \\
&$[45.4 \pm 1.4]$ &$[27.7 \pm 0.9]$ & \\
\hline
\end{tabular}
\end{center}

\vskip 5 truemm
\centerline{\bf Table 2} \par \vskip 3 truemm

Calculated values \cite{5r} of the rapidity distribution $dN/dy$ for
central $Pb$ $Pb \to p -
\overline{p} + X$ and $Pb$ $Pb \to B - \overline{B} + X$ at $\sqrt{s}
= 17.2$~GeV
($k = 1.4$) and central $Au$ $Au \to p - \overline{p} + X$ and $Au$ $Au \to B -
\overline{B} + X$ at $\sqrt{s} = 130$~GeV ($k=2$) and $\sqrt{s} =
200$~GeV ($k=2.2$) .
The centrality has been defined by the number of participant pairs
($n_A = 175$ at all
energies) and $\nu = n/n_A = 4.5$, 5.0 and 5.2 at $\sqrt{s} = 17.2$,
130 and 200~GeV,
respectively. Data are in brackets.

\begin{center}
\begin{tabular}{ccccc}
\hline
$y^*$ &$Pb$ $Pb \to p - \overline{p}$ &$Pb$ $Pb \to B - \overline{B}$
&$Au$ $Au \to p -
\overline{p}$ &$Au$ $Au \to B - \overline{B}$\\
&$\sqrt{s} = 17.2$ GeV &$\sqrt{s} = 17.2$
GeV &$\sqrt{s} = 130$ (200) GeV &$\sqrt{s} = 130$ (200 GeV)\\
\hline
0 &23.0 &58.5 &8.0 (7.4) &20.9 (18.9)  \\
&$[26.7 \pm 3.7]$ &$[67.7 \pm 7.3 ]$ &$[5.6 \pm 0.9 \pm 24\%]$ &  \\
1 &32.3 &79.7 &9.7 (8.7) &22.6 (22.0) \\
&$[34.9 \pm 1.5]$ &$[84.7 \pm 3.5]$ & &\\
1.5 &36.3 &87.0 &12.3 (10.9) &31.5 (27.4) \\
&$[34.4 \pm 1.7]$ &$[80.0 \pm 3.9]$ & &\\
2 &25.3 &57.15 &17.3 (14.3) &43.4 (35.9) \\
&$[24.7 \pm 1.5]$ &$[56.1 \pm 3.1]$ & &\\
\hline
\end{tabular}
\end{center}

\newpage

\par \vskip 1 truecm

\centerline{\bf \large Figure captions}\par \vskip 5 truemm

\noi{\bf Figure 1} : The mechanism of particle production in $e^+e^-$
annihilation. The net of soft gluons and quark loops is only shown here and in
Fig. 6.\par \vskip 5 truemm

  \noi{\bf Figure 2} : One string diagram in
$\overline{p}p$. \par \vskip 5 truemm

\noi{\bf Figure 3} : Dominant two-chain (single
cut Pomeron) contributions to high energy $\pi^+$-proton collisions.
\par \vskip 5 truemm

\noi{\bf Figure 4} : Dominant two-chain
contribution to proton-antiproton collisions at high energies (single cut
Pomeron).\par
\vskip 5 truemm

\noi{\bf Figure 5} : Dominant two-chain diagram describing
multiparticle production in high energy proton-proton collisions (single cut
Pomeron). \par \vskip 5 truemm

\noi{\bf Figure 6} : Single Pomeron exchange and its underlying
cylindrical topology. This is the dominant contribution to proton-proton elastic
scattering at high energies.
\par \vskip 5 truemm

\noi {\bf Figure 7} : Two cut Pomeron (four-chain) diagram for proton-proton
collisions. \par \vskip 5 truemm

\noi {\bf Figure 8} : The values of $dN^{ch}/dy$
per participant for $Pb$ $Pb$ collisions at $\sqrt{s} = 17.3$~GeV computed
\cite{2r} from eq. (2), compared with WA98 data. \par \vskip 5 truemm

\noi {\bf Figure 9} : The values of $dN^{ch}/d\eta_{c.m.}/(0.5\
n_{part})$ for $Au$ $Au$ collisions at $\sqrt{s} = 130$~GeV computed  \cite{2r}
from eq. (2) including shadowing corrections are given by the dark band in
between solid lines. The PHENIX data are also shown (black circles and shaded
area).\par \vskip 5 truemm

\noi {\bf Figure 10} : (a) Calculated values \cite{6r} of $dN/dy$ of $p$ (solid
line) $\overline{p}$ (dashed line), and $p - \overline{p}$ (dotted line) at mid
rapidities, $|y^*| < 0.35$, are plotted as a function of $dN_{h^-}/d\eta$, and
compared with PHENIX data [33]~; (b) same for $\Lambda$ and $\overline{\Lambda}$
compared with preliminary STAR data [34]~; (c) same for $\Xi^-$ and
$\overline{\Xi}^+$ compared to preliminary STAR data [35]~; (d) same for $\Omega$
and $\overline{\Omega}$. \par \vskip 5 truemm

\noi {\bf Figure 11} : Calculated values \cite{6r} of the ratios
$B/n_{part}$ (a) and $\overline{B}/n_{part}$ (b), normalized to the same ratio
for peripheral collisions ($n_{part} = 18$), plotted as a function of
$n_{part}$.\par \vskip 5 truemm

\noi {\bf Figure 12} : The ratio of $J/\psi$ over $DY$ cross-sections
in $Pb$ $Pb$ collisions a 158 GeV/c versus $E_T$ obtained \cite{42r} in the
comovers interaction model with $\sigma_{abs} = 4.5$~mb and $\sigma_{co} =
0.65$~mb. The absolute normalization is 47. The preliminary data are from
[39].\par \vskip
5 truemm

\newpage

\centerline{{\epsfysize18cm \epsfbox{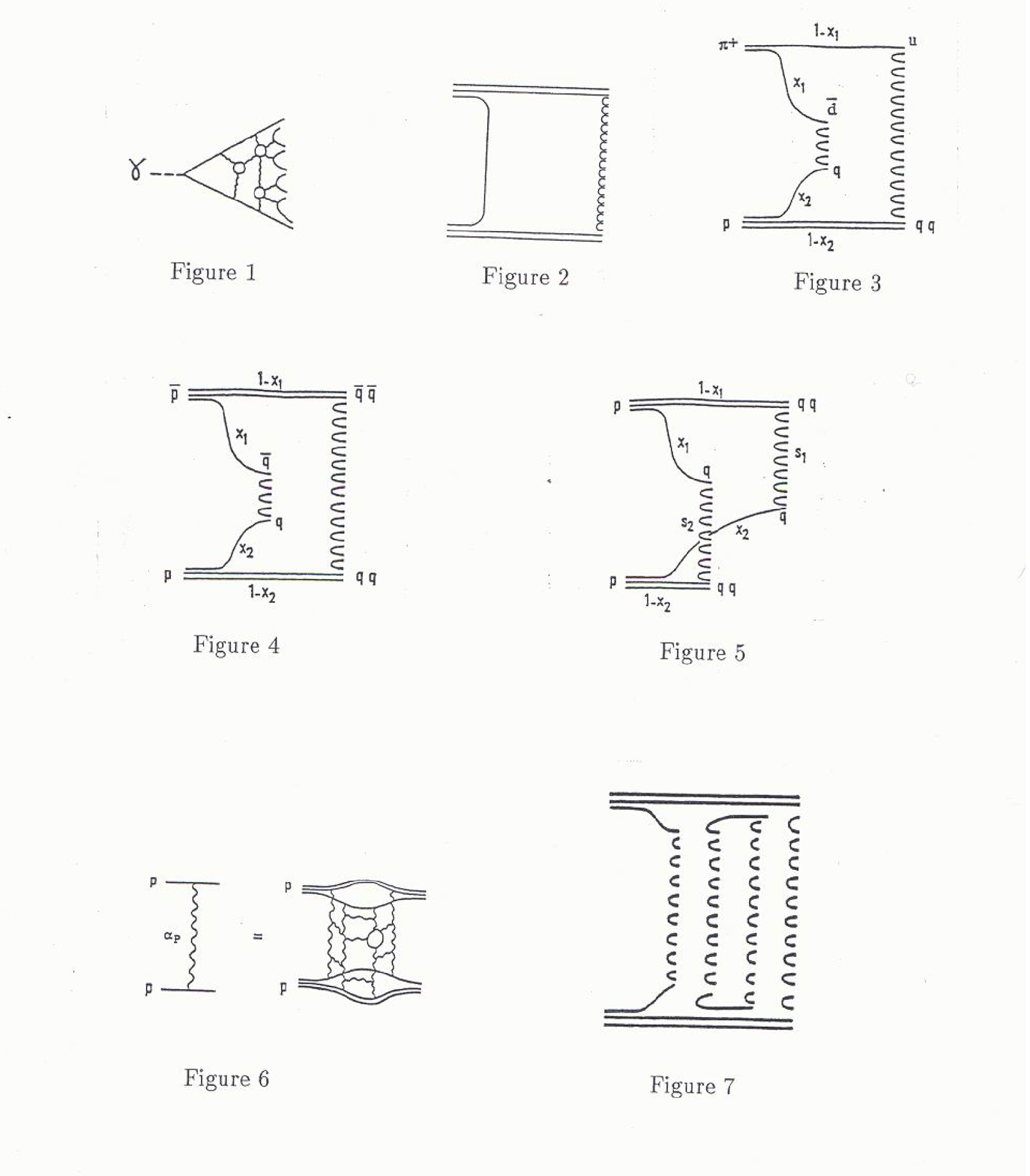}}}

\centerline{{\epsfysize18cm \epsfbox{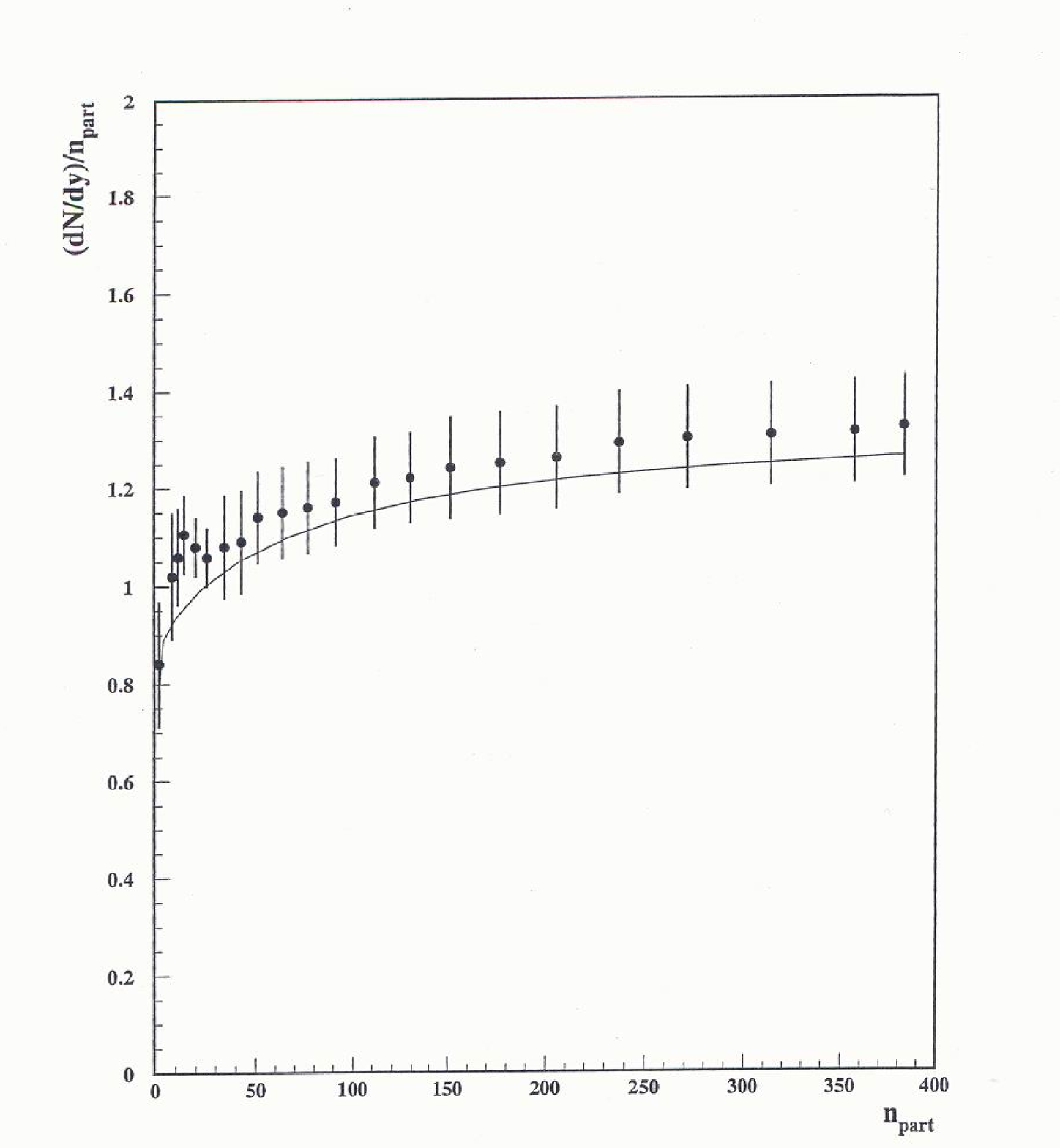}}}
\centerline{Figure 8}

\centerline{{\epsfysize18cm \epsfbox{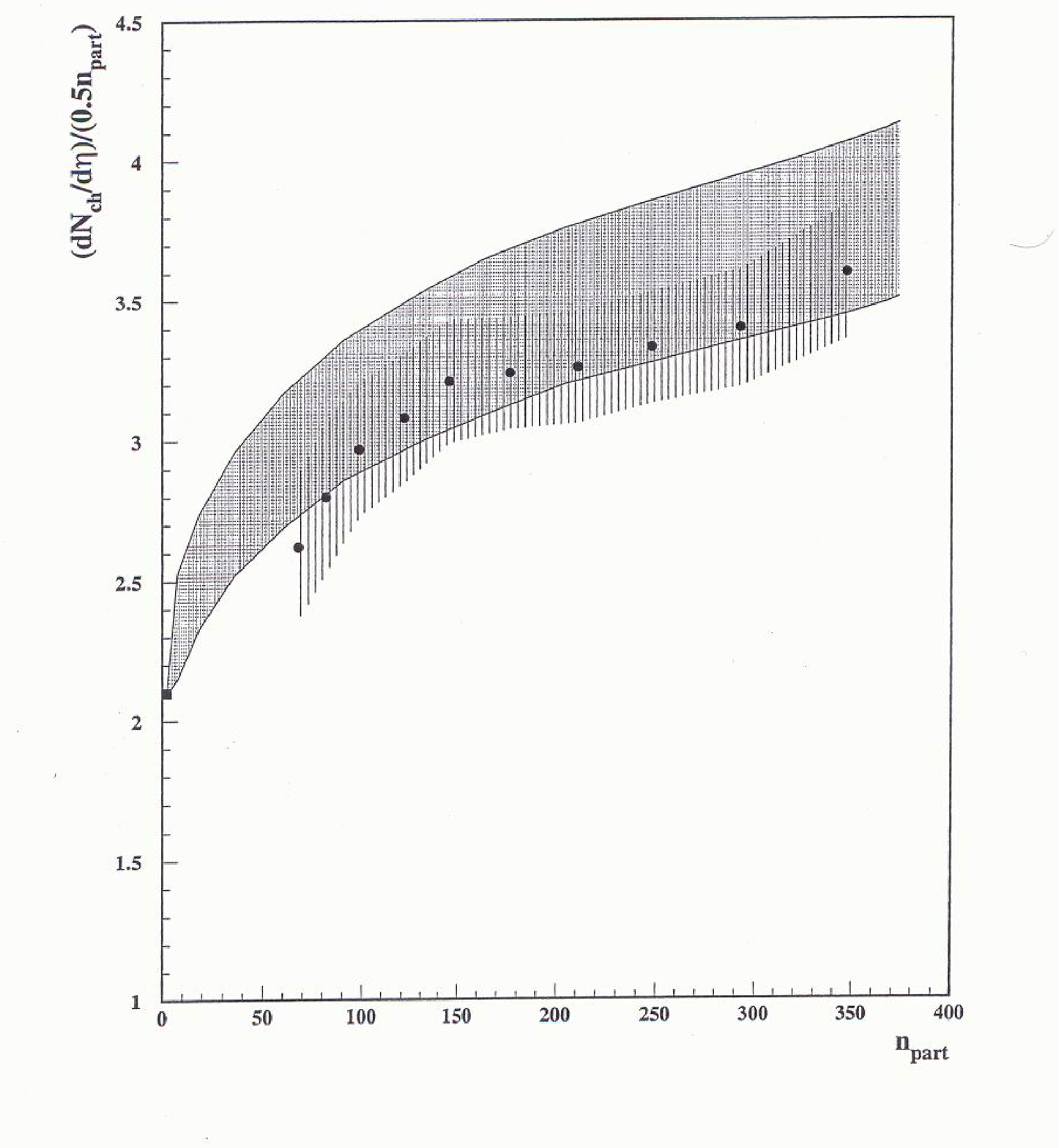}}}
\centerline{Figure 9}

\centerline{{\epsfysize18cm \epsfbox{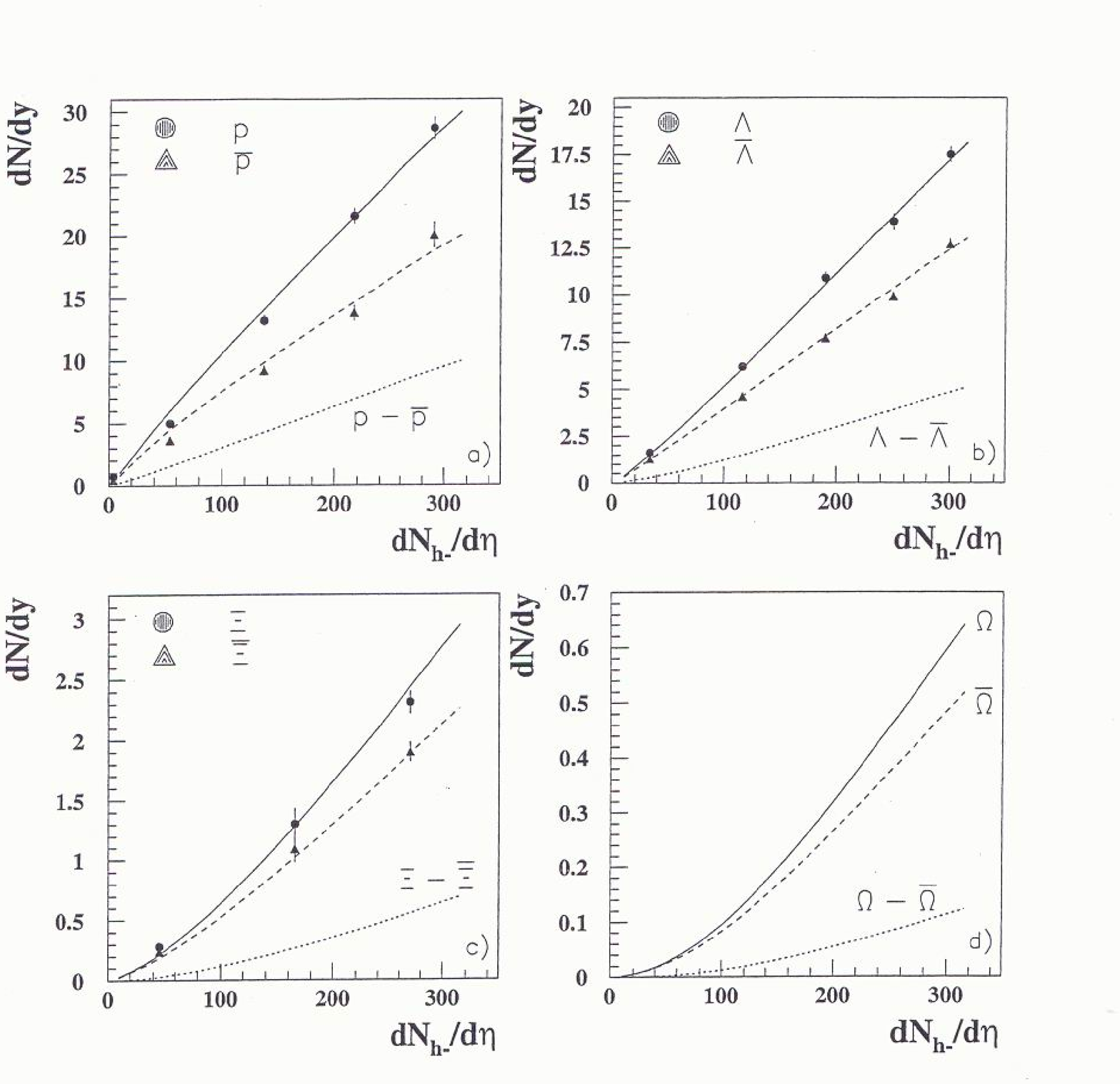}}}
\centerline{Figure 10}

\centerline{{\epsfysize18cm \epsfbox{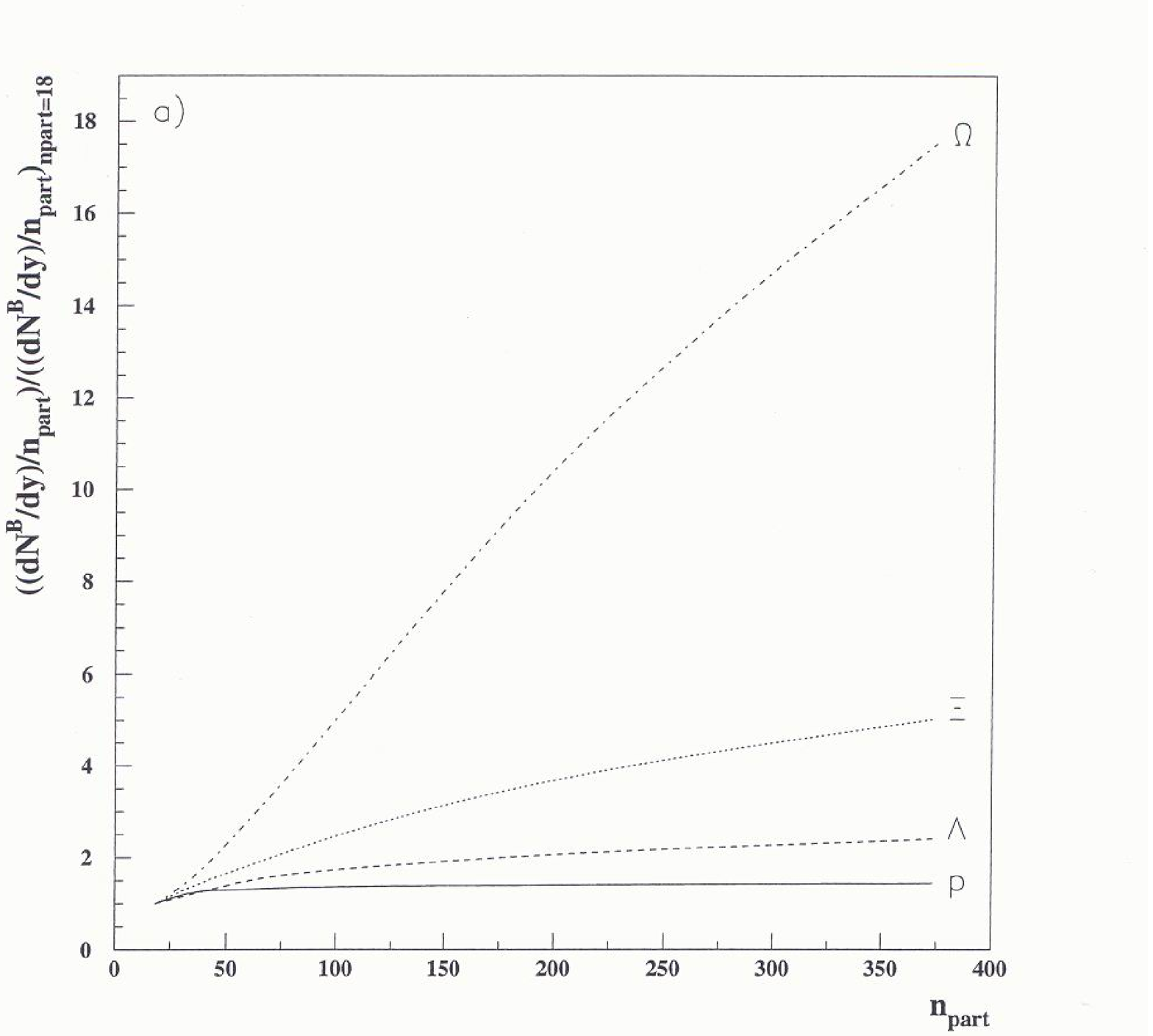}}}
\centerline{Figure 11a}

\centerline{{\epsfysize18cm \epsfbox{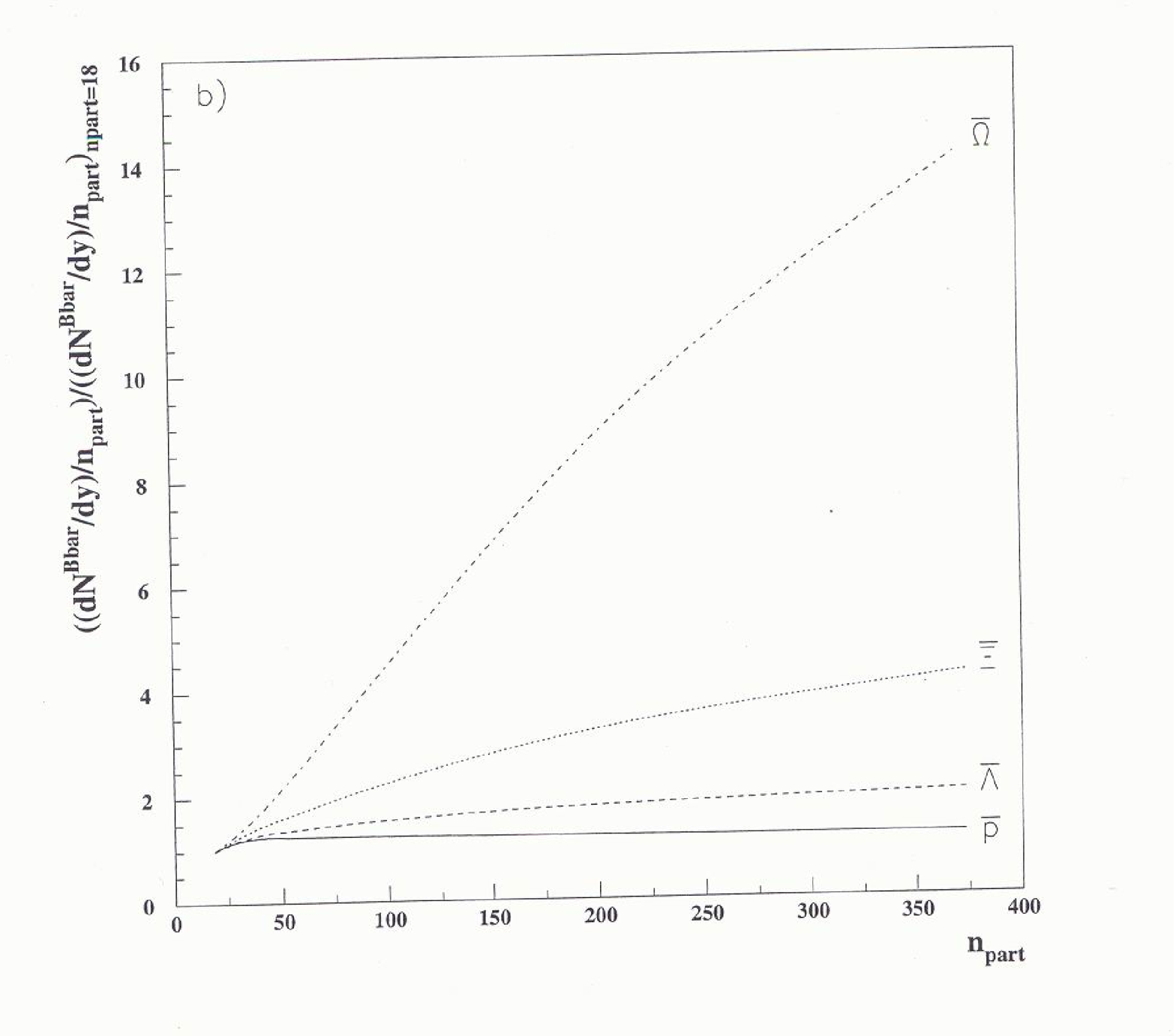}}}
\centerline{Figure 11b}

\centerline{{\epsfysize18cm \epsfbox{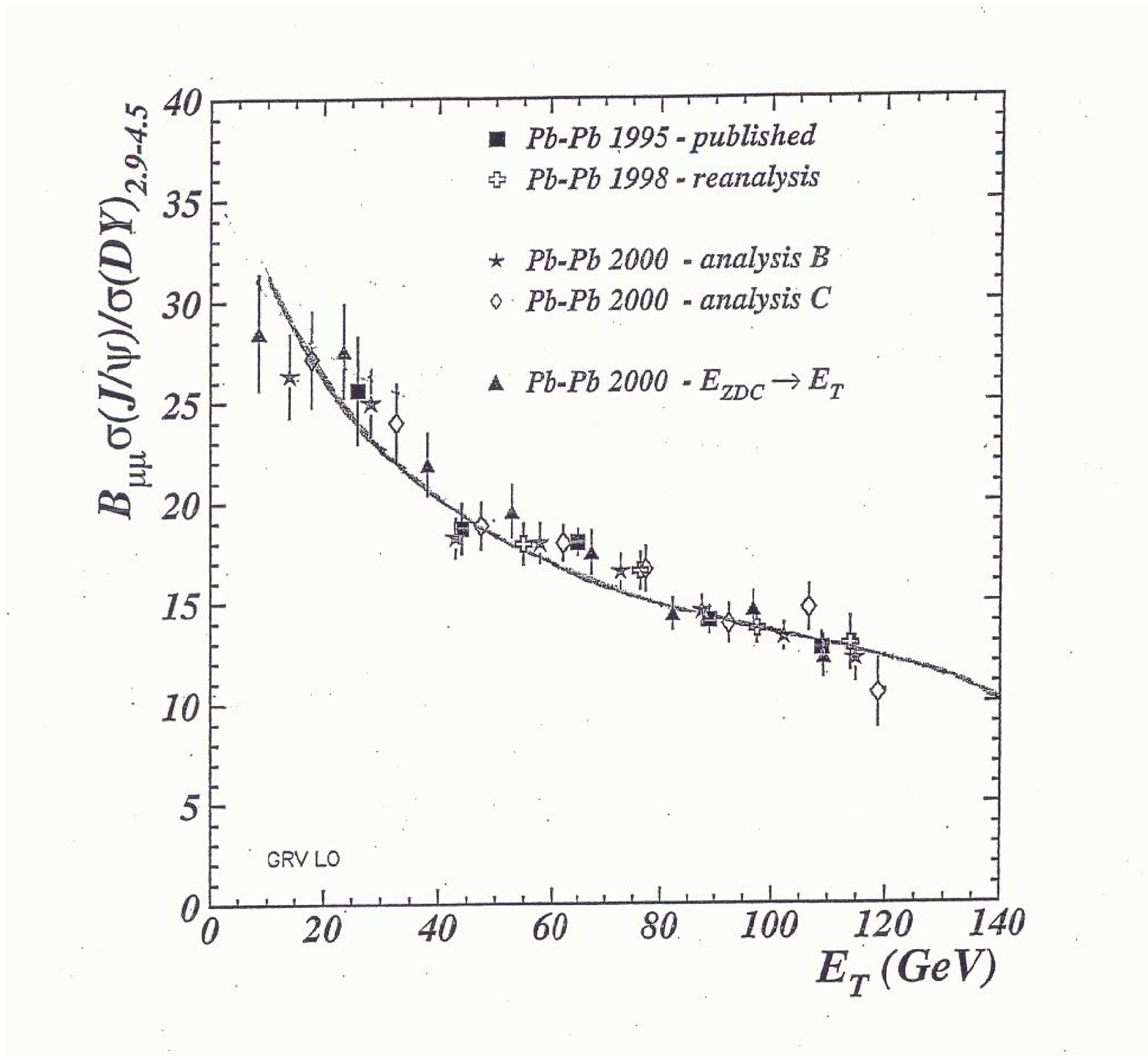}}}
\centerline{Figure 12}


\end{document}